\documentstyle[prd,aps,floats]{revtex}

\begin{document}

\draft
\input epsf
\twocolumn[\hsize\textwidth\columnwidth\hsize\csname
@twocolumnfalse\endcsname

\title{Difficulty of a spinning complex scalar field to be dark
energy}  

\author{S. Kasuya}
\address{Research Center for the Early Universe, University 
  of Tokyo, Bunkyo-ku, Tokyo 113-0033, Japan}

\date{May 23, 2001}

\maketitle

\begin{abstract}
We comment on the possibility of a spinning complex scalar field to be
dark energy. We show that it deforms (almost) completely into a
nontopological soliton state, a Q ball, and the equation of state 
becomes that of the matter or radiation, depending on the fate of the
Q ball. Thus, the spinning complex scalar field is usually very
difficult to play the role of the dark energy. We also show the
general condition that the spinning complex scalar field can
successfully be the dark energy. 
\end{abstract} 

\pacs{PACS numbers: 98.80.Cq, 11.27.+d, 11.30.Fs
      \hspace{46mm} astro-ph/0105408}


\vskip2pc]

\setcounter{footnote}{1}
\renewcommand{\thefootnote}{\fnsymbol{footnote}}

Recent supernovae observation reveals that the expansion of the
universe is accelerating \cite{SN}. In order to explain the fact, we
need the cosmological constant whose fraction to the critical density
is 0.7, for example. However, not only the cosmological constant plays 
a role for the accelerating universe, but also something which has a
negative pressure, called {\it dark energy} in the literature.

Einstein equation tells us that $\ddot{a}/a \propto -(\rho+3p)$. Thus,
when the equation of state is given by $p=w\rho$, $w<-1/3$ should hold 
for the dark energy to be the source of the accelerating universe. For 
example, the cosmological constant has $w=-1$. On the other hand, a 
homogeneous scalar field has $w=(T-V)/(T+V)$, where $T$ and $V$ denote
the kinetic and potential energy of the field, respectively, and it is
possible to have $-1 \le w < -1/3$. Thus, a scalar field, called
quintessence, can be a good candidate for the dark energy.

Recently it was proposed that a spinning complex scalar field with
$U(1)$ potential could play a role of the dark energy, which has some
different feature from quintessence, and dubbed {\it spintessence}
\cite{Spin}. (See also \cite{GuHw}.) They show that a complex scalar
field can be regard as quintessence if the field does not rotate so
much ($\omega \lesssim H$, where $\omega\equiv \dot{\theta}$ is a
phase velocity and $H$ the Hubble parameter). In the opposite case,
when the field rotates (or rapidly spins) in the potential ($\omega
\gg H$), it will be another category of the dark energy,
spintessence. Most important differences of its characters from that
of the quintessence are the smallness of the changes of the equation
of state and the different features in the fluctuation spectra
\cite{Spin}. Although they just comment on the possibility of the
creation of nontopological solitons, the formation of nontopological
solitons, or Q balls, is very generic for a complex field, as we have
shown in the context of the Affleck-Dine baryogenesis
\cite{KK1,KK2,KK3,KK4}. In this letter, we show that a spinning
complex scalar field may be difficult to be the dark energy 
($w<-1/3$), and it will most likely to deform into Q balls, and the
equation of motion becomes that of the matter or radiation depending
on the fate of the Q ball created. We will also show the general
condition of the possibility for the spinning complex scalar field to
be successful dark energy.

A Q ball is a kind of nontopological soliton whose stability is
guaranteed by some charge $Q$ \cite{Coleman}, which is a $U(1)$
charge for a complex field with $U(1)$ symmetric potential. The
condition for the Q ball to exist is that $V(\Phi)/|\Phi|^2$ has the
minimum at $\Phi \ne 0$. This condition is usually met for the
potential of the scalar field whose curvature is negative. $w<-1/3$
leads to the potential flatter than $V \sim |\Phi|$. 

For the scalar field which has negative pressure is generally
unstable, and the fluctuations develop. Moreover, the curvature of the 
effective potential for the scalar field with $w<-1/3$ is negative,
which also leads to spatial instabilities. We first show the
instability band. We write a complex field as 
$\Phi = (\phi e^{i \theta})/\sqrt{2}$, and decompose into homogeneous   
parts and fluctuations: $\phi \rightarrow \phi +\delta\phi$ and
$\theta \rightarrow \theta + \delta\theta$. Assuming that the gravity
effects are weak, which is a good approximation here, we obtain the
equation of motion of the $\Phi$ field as \cite{KuSh,EnMc,KK1,KK2,KK4} 
\begin{eqnarray}
    \label{phi-eq}
    \ddot{\phi} + 3H\dot{\phi} - \dot{\theta}^2\phi
      + V'(\phi) & = & 0, \\
    \label{theta-eq}
    \phi\ddot{\theta} + 3H\phi\dot{\theta} 
      + 2\dot{\phi}\dot{\theta} & = & 0,
\end{eqnarray}
for the homogeneous mode, and
\begin{eqnarray}
    \label{eom-fl}
    \delta\ddot{\phi} + 3H\delta\dot{\phi}
     - 2\dot{\theta}\phi\delta\dot{\theta} - \dot{\theta}^2\delta\phi
     -\frac{\nabla^2}{a^2}\delta\phi + V''(\phi)\delta\phi & = & 0, 
     \\
     \phi\delta\ddot{\theta} 
       + 3H \phi\delta\dot{\theta}
       + 2(\dot{\phi}\delta\dot{\theta} 
           +\dot{\theta}\delta\dot{\theta})
       -2\frac{\dot{\phi}}{\phi}\dot{\theta}\delta\phi
       -\phi\frac{\nabla^2}{a^2}\delta\theta & = & 0
\end{eqnarray}
for fluctuations. Equation (\ref{theta-eq}) represents the
conservation of the charge (or number) within the physical volume:
$Q=\dot{\theta}\phi^2a^3=const.$ 

We seek for the solutions in the form
\begin{equation}
    \delta\phi = \delta\phi_0 e^{\alpha(t)+ikx}, \qquad
    \delta\theta = \delta\theta_0 e^{\alpha(t)+ikx}.
\end{equation}
If $\alpha$ is real and positive, these fluctuations grow
exponentially, and go nonlinear to form Q balls. Inserting these forms 
into Eqs.(\ref{eom-fl}), we get the following condition for nontrivial 
$\delta\phi_0$ and $\delta\theta_0$, 
\begin{eqnarray}
   & & \left(\ddot{\alpha}+\dot{\alpha}^2+3H\dot{\alpha}
              +\frac{k^2}{a^2}+V''-\dot{\theta}^2\right)
   \nonumber \\
   & & \hspace*{5mm}
   \times \left(\ddot{\alpha}+\dot{\alpha}^2+3H\dot{\alpha} 
                +\frac{k^2}{a^2}
                +\frac{2\dot{\phi}\dot{\alpha}}{\phi}\right)
   + 4\dot{\theta}^2\dot{\alpha}^2 = 0.
\end{eqnarray}
This equation can be simplified to be
\begin{equation}
    \label{det}
    \dot{\alpha}^4 
     + \left( 2\frac{k^2}{a^2} + V'' + 3\dot{\theta}^2  \right)
                \dot{\alpha}^2  
     + \left( \frac{k^2}{a^2} + V'' - \dot{\theta}^2 \right) 
            \frac{k^2}{a^2} = 0,
\end{equation}
where we assume that cosmological expansion is negligible, $H\sim 0$,
so that the orbit of the field in the potential is circular: 
$\phi \sim const.$ We also assume that
$\ddot{\alpha}\ll\dot{\alpha}^2$. 

In order for $\alpha$ to be real and positive, we must have the last
term of Eq.(\ref{det}) to be negative, so that the instability band
for the fluctuations is 
\begin{equation}
    0 < \frac{k^2}{a^2} < \dot{\theta}^2 -V''.
\end{equation}
Notice that the instability band always exist, since the curvature of
the potential is negative for $w<-1/3$.

Q balls are produced with the typical size $\sim k_{res}^{-1}$, where
$k_{res}$ denote the most amplified mode in the instability band. This
process generally takes place nonadiabatically as we showed in
Refs.~\cite{KK1,KK2,KK4}. Once the Q balls are produced, they act like 
a (dark) matter, so the energy density evolves as $a^{-3}$. We also
show that (almost) all the charges of the field are absorbed into the
produced Q balls, so that there is no homogeneous field left to be a
dark energy. For the later fate of the Q ball, the most important
feature of the Q ball is its mass in the function of the charge. It
can be written as 
\begin{equation}
    M_Q \sim m_{\phi,eff} Q^p,
\end{equation}
where $m_{\phi,eff} \sim |V''|$ at the Q-ball formation time, and 
$3/4 \le p \le 1$, depending on the shape of the potential. For
example, $p=3/4$ for the flat potential \cite{Dvali}. The Q ball is
stable against the decay into some other particles, if the mass per
unit charge is smaller than the mass of those particles:
\begin{equation}
    \frac{M_Q}{Q} \sim m_{\phi,eff} Q^{-(1-p)}< \frac{m_{decay}}{q}, 
\end{equation}
where $m_{decay}$ is the smallest mass of the particles which has the
same $U(1)$ charge as the complex field $\Phi$, and $q$ is the charge
carried by this particle. If this condition
holds, Q balls with large enough charge to be stable against the decay 
into other particles can exist in the present
universe, and it can be a dark matter if their energy density has a
crucial fraction to the critical density. For the smaller charge, Q
balls  disappear through the charge evaporation from the Q-ball
surface, and $\Phi$-particles may becomes a component of the
radiation. On the other hand, they decay into (lighter) particles, and 
those particles contribute to the radiation, and may be some fraction
of the (dark) matter depending on their masses. Therefore, the
energy density evolves as $a^{-4}$ or $a^{-3}$. In any case, they
cannot contribute to the dark energy, whose energy density evolves as
$a^{-r}$ with $r<3$.

One may wonder how general the Q ball formation occurs for the
spinning complex scalar field dark energy. We will seek for the
possibility of the potential which does not lead to Q-ball formation,
while the field still acts as the dark energy. 
\footnote{
We thank M. Kamionkowski and T. Yanagida for suggesting this
possibility.}
The kinetic energy is
written as 
\begin{equation}
    |\dot{\Phi}|^2 = \frac{1}{2}(\dot{\phi}^2+\dot{\theta}^2\phi^2)
    \simeq \frac{1}{2}\dot{\theta}^2\phi^2,
\end{equation}
where we assume $\dot{\theta}\gg H$ and $\dot{\phi} \approx 0$ in the
last equality. In order for the accelerating expansion, we need 
$w < -1/3$, so 
\begin{equation}
    \dot{\theta}^2 < \frac{V(\phi)}{\phi^2},
\end{equation}
where we use $V(\Phi)=V(\phi)$ because of the $U(1)$-symmetric
potential. From Eq.(\ref{phi-eq}) with $H\approx 0$, we have 
$\phi\dot{\theta}^2\simeq V'(\phi)$, so that
\begin{equation}
    \label{dark-cond}
    \ddot{a} > 0 \quad \rightarrow \quad \phi V'(\phi)<V(\phi).
\end{equation}
Rewriting this condition in terms of $f(\phi)\equiv V(\phi)/\phi^2$,
we obtain
\begin{equation}
    \label{dark-cond-2}
    f(\phi) + \phi f'(\phi) < 0.
\end{equation}
On the other hand, the condition of the existence of the Q ball is
that the function $f(\phi)$ has the minimum value $f_{min}$ at 
nonzero $\phi$:
\begin{equation}
    \label{Q-cond}
    f=f_{min} \quad {\rm at} \quad \phi=\phi_* \ne 0.
\end{equation}

Suppose the situation that the Q-ball condition
(\ref{Q-cond}) does not hold. We will find whether the dark-energy
condition (\ref{dark-cond}), or, equivalently, (\ref{dark-cond-2}),
can be satisfied in this case. The function $f(\phi)$ which violates
the Q-ball condition can be divided into two types: (a) monotonously
increasing function, and (b) the function which has an extremum
(or some extrema), but the minimum of the function is achieved at
$\phi=0$. For the first type, $f'>0$ is always satisfied, so that the 
dark-energy condition (\ref{dark-cond-2}) cannot hold. This implies
that the spinning complex scalar field in the type (a) potential
cannot be the dark energy of the universe. 

For example of this type, we consider the superposition of power-law
potential, $V(\phi) = \sum_i A_i \phi^{q_i}$. ($q_i$'s cannot be all
negative, since the condition for the complex field to spin rapidly 
in the potential is $(\phi^3V'(\phi))'>0$ \cite{Spin}, assuming
$V(0)=0$.) The dark-energy condition is expressed as 
\begin{equation}
    \sum_i (q_i-1)A_i\phi^{q_i} < 0,
\end{equation}
which can be satisfied if some $i$ exist such that $q_i<1$. On the
other hand, $f(\phi)=\sum_i A_i\phi^{q_i-2}$. It is easily seen that
$f=f_{min}$ is achieved at $\phi=\infty$ in the case that $q_i<2$ for
all $i$. If some $j$ exist such that $q_j>2$, 
$f'(\phi)=\sum_i (q_i-2)A_i\phi_{q_i-3}=0$ is satisfied at $\phi_*$,
which is determined by
\begin{equation}
    \sum_j (q_j-2)A_j\phi_*^{q_j-3}=\sum_k (2-q_k)A_k\phi_*^{q_k-3}, 
\end{equation}
where $q_j >2$, while $q_k<2$. In any case, the minimum of the
function $f(\phi)$ is achieved at nonzero $\phi$, which leads to the 
Q-ball formation.

In the second category (b), it is possible for the field to be dark
energy in general in some range of the field amplitude. During the
field to be the dark energy, where the field feels negative pressure, 
fluctuations develops very effectively, and the amplitude of the field
may decreases more rapidly. This may lessen the effects for the dark
energy. The examples of this type of the effective potential are 
$V(\Phi)=|\Phi|^2((|\Phi|^2)^{-r/2}+A)\exp(-1/|\Phi|^2)$, where
$1<r<2$ and $A> 0$, or $V(\Phi)=(|\Phi|^2-B)\exp(-|\Phi|^2/M^2)+B$,
where $B> 0$ and $M$ is some mass scale. We show plot $f(\phi)$ of
these examples in Fig.1. It seem somewhat difficult to obtain such
form of the potentials, but not impossible. 

\begin{figure}[t!]
\centering
\hspace*{-7mm}
\leavevmode\epsfysize=6cm \epsfbox{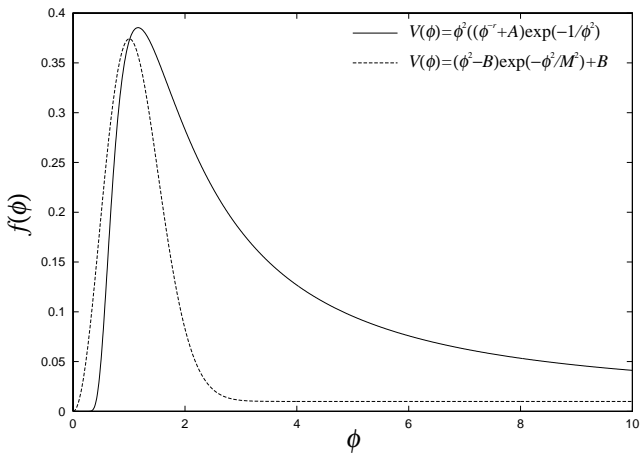}\\[2mm]
\caption[fig1]{\label{fig1} 
Examples of $f(\phi)$ for the successful dark energy spinning complex
scalar field.}
\end{figure}

Finally, we must mention the fate of the complex scalar field which is
just oscillating along the radial direction in its potential. This is
the very situation for the so-called oscillating inflation \cite{OSC}. 
Even in this case, the Q-ball formation takes place naturally: both Q
and anti-Q balls with the opposite sign of charges of the same order
of magnitude are created \cite{KK1,KK2,KK4}. Therefore, the universe
cannot expand in the accelerated rate because of the same reason
mentioned above.

In summary, we have shown that a spinning complex scalar field feels
spatial instabilities, and deforms into Q balls very generally. Once
the Q balls are produced, it acts like a (dark) matter, and the
equation of state becomes $p=0$. The most important fact is that
(almost) all the charges of the complex field are absorbed into the
produced Q balls, so that there is no (homogeneous) field left to be a
dark energy. Concerned with the later fate of the Q balls, it depends
on the shape of the potential, and they can be stable to be the dark
matter, or decay into other particles to be the radiation. (Complete
evaporation of Q balls also leads to the radiation component.) In
either case, the energy of the complex field are altered into the form
of matter or radiation, and it cannot play a role for the source of
the accelerating universe, the dark energy. Of course, if the rotation 
speed of the field is very slow ($\omega \ll H$), the field slowly
rolls down to the minimum of the potential, and acts as the
quintessence, as mentioned in Ref.\cite{Spin}, and can be the dark
energy. 

We have also shown that there is some possibility for the spinning
complex field to be the dark energy. It may realized if the potential
of the field has somewhat unusual forms such that $f(\phi)$ should
have an extremum at nonzero $\phi$ and its minimum at $\phi=0$.

Concerned with symmetries other than global $U(1)$, it is known that
the Q-ball type soliton exists in many cases, such as for nonabelian
symmetries \cite{nonAbelian}, gauged $U(1)$ \cite{GaugedQ}, etc. 
Therefore, if the fluctuations of the field develop enough, these Q
balls might be created in very similar manner, and their energy
density evolves as the matters.

The author is grateful to M. Kawasaki for useful discussions.

\end{document}